\documentclass[%
 reprint,
frontmatterverbose, 
showpacs,preprintnumbers,
nofootinbib,
 amsmath,amssymb,
 aps,
floatfix,
]{revtex4-1}

\usepackage{graphicx}
\usepackage{dcolumn}
\usepackage{bm}

\usepackage{graphicx}
\usepackage{xcolor}
\usepackage{tabularx}
\usepackage{fancybox}
\usepackage{slashed}
\usepackage{amsfonts}
\usepackage{empheq}
\usepackage{bbold}
\usepackage{amssymb}
\usepackage{mathtools}
\usepackage[latin1]{inputenc}
\usepackage[T1]{fontenc}
\usepackage[francais]{babel}
\usepackage{upgreek}
\usepackage{amsmath}
\usepackage{slashed}
\usepackage{tikz}
\usepackage{graphicx}
\usepackage{pgfplots}
\usepackage{color}
\usepackage{rotating}
\usepackage{gensymb}
\usepackage{simplewick}
\usepackage{enumerate}
\usepackage{ytableau}
\usepackage{youngtab}
\usepackage{multirow}
\usepackage{tabularx}
\usetikzlibrary{quotes,angles,snakes}
\usetikzlibrary{trees}
\usetikzlibrary{decorations.pathmorphing}
\usetikzlibrary{decorations.markings}
\usetikzlibrary{arrows,shapes,positioning}
\usetikzlibrary{decorations.markings}
\tikzstyle arrowstyle=[scale=1]
\tikzstyle directed=[postaction={decorate,decoration={markings,
    mark=at position .65 with {\arrow[arrowstyle]{stealth}}}}]
\tikzstyle reverse directed=[postaction={decorate,decoration={markings,
    mark=at position .65 with {\arrowreversed[arrowstyle]{stealth};}}}]

\newcommand{\RNum}[1]{\uppercase\expandafter{\romannumeral #1\relax}}

\newcommand{\beq}{\begin{equation}}
\newcommand{\eeq}{\end{equation}}
\newcommand{\bea}{\begin{eqnarray}}
\newcommand{\eea}{\end{eqnarray}}
\usepackage{ulem}

\usepackage{xcolor,colortbl}
\definecolor{Blue}{RGB}{140,165,195}
\definecolor{Purple}{RGB}{255,145,145}
\definecolor{bluc}{cmyk}{1,1,0,0.1}
\definecolor{rossoCP3}{cmyk}{0,.88,.77,.40}
\definecolor{rosso}{cmyk}{0,1,1,0.4}
\definecolor{rossos}{cmyk}{0,1,1,0.55}
\definecolor{rossoc}{cmyk}{0,1,1,0.2}
\definecolor{verdes}{cmyk}{0.92,0,0.59,0.4}

\newcommand{\SU}{\mathrm{SU}}

\newcommand{\UU}{\mathrm{U}}

\usepackage{color}
\usepackage{ulem}

\begin{document}

\preprint{CP3-Origins-2019-34 DNRF90}

\title{Tumbling to the Top}

\author{Giacomo Cacciapaglia$^{1,2}$}
\author{Shahram Vatani$^{1,2}$}
\author{Zhi-Wei Wang$^{3,4}$}
\affiliation{$^{1}$Institut de Physique des 2 Infinis (IP2I), CNRS/IN2P3, UMR5822, F-69622 Villeurbanne, France\\
$^{2}$Universit\'e de Lyon, France; Universit\'e Claude Bernard Lyon 1, Lyon, France\\
$^{3}$$\rm{CP}^3$-Origins, University of Southern Denmark, Campusvej 55 5230 Odense M, Denmark\\
$^{4}$Department of Physics, University of Waterloo, Waterloo, On N2L 3G1, Canada
}

\begin{abstract}
We propose a new mechanism to generate the couplings of the top quark to a composite Goldstone Higgs, which we dub {\it Global Extended Technicolor} (GETC). Top, techni-fermions and spin-1 mediators arise as bound states of a tumbling chiral gauge theory. 
We propose a simple model based on a tumbling $\SU(4)$ dynamics, which predicts partial compositeness for the top and an $\SU(3)$ model with 6 Dirac flavours for the low-energy composite Higgs. Neutron decay bounds require the new dynamics to confine above $10^{6\div 7}$~TeV. A simple extension of the basic model allows to generate partial compositeness also for bottom and tau, and a walking dynamics from an intermediate $\SU(3)$ theory with 10 flavours (6 light + 4 heavy), whose dynamics can be studied on the Lattice.  
\end{abstract}
\maketitle

The idea that the electroweak symmetry may be broken by a confining strong dynamics is as old as the Standard Model (SM) itself~\cite{Weinberg:1975gm,Susskind:1978ms}. While the first attempts were essentially Higgsless, it was later realised that a Higgs-like light scalar can easily arise as a pseudo-Nambu-Goldstone Boson (pNGB) of an extended symmetry breaking~\cite{Kaplan:1983fs}. This paradigm recently received new life~\cite{Contino:2003ve} after the discovery of holography, which established a link between theories in extra dimensions and near conformal dynamics in four dimensions~\cite{Maldacena:1997re}. Most of the recent literature has been focusing on generating a little hierarchy between the electroweak and the confining scales, required by avoiding electroweak precision bounds and preserving the SM-like couplings of the Higgs candidate (see, for instance, Refs~\cite{Bellazzini:2014yua,Panico:2015jxa} and references therein). While this requires some level of tuning, it can be easily engineered in realistic models.

The major challenge facing this class of theories is the origin of masses for the SM fermions (we recall that the coupling to gauge bosons are generated by partly gauging the global symmetries of the strong sector), and in particular understanding why the top has such a large coupling. In an effective approach, based on the holographic extra-dimensional dual, an explanation is given in terms of anomalous dimensions of the operators that couple to the elementary top quark fields~\cite{Contino:2004vy,Cacciapaglia:2008bi}. However, there is no proof that such a conformal field theory exists. Another approach to the problem would consist on proposing an underlying gauge-fermion description for the confining strong dynamics. In this context, the couplings of the SM fermions typically appear as four-fermion interactions, suppressed by a new scale external to the strong dynamics. 
The classic solution, dubbed Extended Technicolor (ETC)~\cite{Eichten:1979ah}, consists in extending the confining Technicolor (TC) gauge interactions and part of the SM ones into a larger group. Once broken, the massive ETC gauge bosons generate the required interactions. However, this approach has two main drawbacks: it is hard to break the ETC gauge group in the desired way, and it is extremely challenging to generate the flavour structure without violating bounds on flavour observables~\cite{Dimopoulos:1980fj}.

In this letter, we propose a new mechanism where the mediators are spin-1 states that are not associated to any gauge interactions. In our proposal, the TC gauge interactions and the SM ones are unified in a {\it global} symmetry group, generated by a new high-scale confining dynamics. We, therefore, dub this mechanism {\it ``Global Extended Technicolor''} (GETC). The main advantage of our proposal is that the extended symmetry is either broken dynamically or not broken at all, thus avoiding the challenges of gauge unification of the classic ETC mechanism. 
Upon confinement, the new GETC dynamics needs to generate spin-1 resonances $\rho^\mu_{\rm GETC}$ (the mediators) and massless baryons $\mathcal{B}$ that contain both the SM fermions and the techni-fermions charged under TC gauge interactions. Once the massive $\rho^\mu_{\rm GETC}$ are integrated out, four-fermion interactions are generated via the $g_\rho\ \rho^\mu_{\rm GETC}\ \bar{\mathcal{B}} \gamma_\mu \mathcal{B}$ coupling and suppressed by $g_\rho/M_\rho \approx 1/f_{\rm GETC}$. No light pNGBs are needed, thus the global symmetry of the GETC dynamics can remain unbroken.
The simplest choice for the GETC dynamics is a chiral gauge theory. In fact, vector-like gauge theories have been proven to always confine and break the global symmetries~\cite{Witten:1983tx,Kosower:1984aw}. Confining chiral gauge theories, on the other hand, have a much richer phase space that we briefly review below.

\begin{table}[t!]
\begin{tabular}{|c|c|c|c|}
\hline
\multicolumn{2}{|c|}{chiral family} & global sym. & massless ferm. \\
\hline
$\xi_A:~\tiny\yng(1,1)$ & $\xi_{\overline{F}}:$~$(N-4) \times \overline{\tiny\yng(1)}$ & $\SU(N-4) \times \UU(1)$ &$\mathcal{B_S}:$~$\tiny\yng(2)$ \\
\hline
$\xi_S:~\tiny\yng(2)$ & $\xi_{\overline{F}}:$~$(N+4) \times \overline{\tiny\yng(1)}$ & $\SU(N+4) \times \UU(1)$ &$\mathcal{B_A}:$~$\tiny\yng(1,1)$ \\
\hline
 \end{tabular} 
\caption{Two basic prototypes of chiral gauge theories: the Bars-Yankielowicz theory (bottom row) and the generalised Georgi-Glashow theory (upper row) are presented. $\xi_A$, $\xi_S$ and $\xi_{\overline{F}}$ denote Weyl fermions in the two-index antisymmetric, symmetric and conjugate-fundamental representations of $\SU(N)$ respectively.}
\label{chiral family} 
\end{table}

The two simplest prototype confining chiral gauge theories are shown in Table~\ref{chiral family}. They are based on a gauged $\SU(N)$ group, with one Weyl spinor in the symmetric representation and $(N+4)$ ones in the conjugate fundamental (Bars-Yankelowicz theory~\cite{Bars:1981se,Appelquist:1999vs}) or one Weyl spinor in the anti-symmetric and $(N-4)$ ones in the conjugate fundamental (generalised Georgi-Glashow theory~\cite{Appelquist:1999vs}). In both cases the gauge coupling is asymptotically free, and thus the gauge interactions are expected to confine in the Infra-Red (IR).
It was firstly proposed in Ref.~\cite{Raby:1979my} that dynamical gauge symmetry breaking may happen sequentially and break the original gauge group into smaller and smaller subgroups. This ``tumbling'' mechanism is an interesting way to dynamically generate hierarchical scales~\cite{Raby:1979my,Georgi:1981mh}, and it has been studied as an explanation of the mass hierarchies of the quarks and leptons 
or as a self-breaking Grand Unification Theory. Later `t Hooft provided an alternative confining picture where the fermions confine without non-zero condensates, and massless composite fermions appear in order to match the global anomalies of the un-broken global symmetries in the confined and unconfined phases~\cite{tHooft:1979rat}. This picture was further supported in Ref.~\cite{Eichten:1981mu} where the most attractive channel (MAC) analysis of tumbling theory is debated. 
Inspired by the lattice work of Ref.~\cite{Fradkin:1978dv}, which shows that there is no sharp boundary between the Higgs (symmetry breaking) phase and confining one for lattice gauge theories with scalars, in Ref.~\cite{Dimopoulos:1980hn} the idea of ``complementarity'' was proposed. 
This idea is based on the observation that the low energy spectrum of the tumbling phase and of the confining one may have exactly the same properties, suggesting that the two describe the same physical phase of the theory in the IR. This property is featured by the generalised Georgi-Glashow theories shown in Table~\ref{chiral family}.
A more recent method that may allow to choose the correct IR phase of the theory is based on free energy analysis. In fact, the IR free energy of a strongly interacting theory should always be lower than that of the UV free theory, as shown in Refs~\cite{Appelquist:1999hr,Appelquist:1999vs}. In Ref.~\cite{Appelquist:2000qg}, this criterium has been extended to proposing that the correct IR vacuum should have the lowest free energy. Interestingly, this would suggest that for both theories in Table~\ref{chiral family} the IR vacuum  corresponds to the confined theory without chiral condensates.

Inspired by the above insights, in this work we propose to construct the first GETC model based on a confining chiral gauge theory. To show that this mechanism can indeed work, we focus on a toy-model where only the top mass is generated. As mentioned above, this is justified by the fact that the top mass is the most challenging one to explain, while masses for lighter states can be generated by other dynamics at a much higher scale. Note also that, taking the engineering dimension of the four-fermion interactions, the top Yukawa can be estimated to
\beq \label{eq:ytop}
y_{\rm top} \approx \left( \frac{\Lambda_{TC}}{f_{\rm GETC}} \right)^2 \approx 1\,,
\eeq
where $\Lambda_{\rm TC} \approx 4 \pi f$ is the TC condensation scale. Thus, it would be enough to have $f_{\rm GETC} \approx \Lambda_{\rm TC}$, so that the confinement scale of the GETC dynamics is one order of magnitude larger than that of the TC interactions ($\Lambda_{\rm GETC} \approx 4 \pi f_{\rm GETC} \approx 4 \pi \Lambda_{\rm TC}$). A larger hierarchy between the GETC and TC scales can also be made phenomenologically acceptable if the theory approaches an IR fixed point in between~\cite{Holdom:1981rm}. 
We will now present the explicit model, explaining the reasons for this choice and its uniqueness while we describe its properties.

The model we propose is based on a confining chiral gauge symmetry $\mathcal{G}_{\rm GETC} = \SU(4)$, with one Weyl fermion in the symmetric and 8 Weyl fermions in the conjugate fundamental of the gauge interactions. In the confining phase, the anomaly matching requires that one baryon in the antisymmetric representation of the global $\SU(8)$ remains massless~\cite{Bars:1981se}. The main reason for this choice is that the massless baryon must contain the quark doublet $q_L$, which transforms as a fundamental of QCD colour, and the right-handed top $t_R^c$, which transforms as the anti-fundamental of QCD colour.~\footnote{Here all Weyl fermions are considered left-handed, so we will use charge-conjugate spinors, like $t_R^c$, to describe right-handed fields.}
The fermion content of the theory (including the third generation of SM fermions) is summarised in Table~\ref{fundamental theory}, where we list he quantum numbers under $\mathcal{G}_{\rm GETC}$, the Technicolor group $\mathcal{G}_{\rm TC} = \SU(3)$, and the SM gauge interactions. 

\begin{table}
\begin{tabular}{|c|c|c|c|c|c||c|}
\hline
 & $\SU(4)_{\rm GETC}$ &  $\SU(3)_{\rm TC}$ & $\SU(3)_c$ & $\SU(2)_L$ & $\UU(1)_Y$ & $\UU(1)_{\rm B-L}$ \\
\hline
$\xi_S$ & $\tiny\yng(2)$ & $1$ & $1$ & $1$ &  $0$&$0$\\
$\xi_{\bar{F}}^1$ & $\overline{\tiny\yng(1)}$ & $1$ & $\tiny\yng(1)$ & $1$ & $-1/3$&$-1/6$\\
$\xi_{\bar{F}}^2$ & $\overline{\tiny\yng(1)}$ & $\tiny\yng(1)$ & $1$ & $1$ & $0$&$-1/6$\\
$\xi_{\bar{F}}^3$ & $\overline{\tiny\yng(1)}$ & $1$ & $1$ & $\tiny\yng(1)$ & $1/2$&$1/2$\\
\hline
$b_R^c$ & $1$ & $1$ & $\overline{\tiny\yng(1)}$ & $1$ &  $1/3$&$1/3$\\
$l_L$ & $1$ & $1$ & $1$ & $\tiny\yng(1)$ &  $-1/2$&$-1$\\
$Q_R^{c}$ & $1$ & $\overline{\tiny\yng(1)}$ & $1$ & $\overline{\tiny\yng(1)}$ & $-1/2$&$1/3$\\
$N_L$ & $1$ & $\tiny\yng(1)$ & $1$ & $1$ & $0$&$1/3$\\
$X_R^c$ & $1$ & $\overline{\tiny\yng(1)}$ & $\overline{\tiny\yng(1)}$ & $1$ & $1/3$&$-1/3$\\
\hline
 \end{tabular} 
 \caption{Fermion content in the minimal GETC model, based on a gauged $\SU(4)$ Bars-Yankielowicz theory (top block). The elementary fermions in the lower block are added to cancel gauge anomalies and include the third generation right-handed bottom and lepton doublet. The last column contains the charge assignment under a global $\UU(1)_{\rm B-L}$ charge preserved in the model.}
 \label{fundamental theory}
 \end{table}

The quantum numbers of the $\xi_{\bar{F}}$ fermions are chosen in order to embed $\SU(3)_{\rm TC}$, $\SU(3)_c$ and $\SU(2)_L$ in the global symmetry (thus $8 = 3 + 3 + 2$ is the minimal number of $\xi_{\bar{F}}$, fixing $N=4$ for the GETC gauge group). The hypercharges are fixed by the requirement of the massless baryon to contain the SM quark doublet $q_L$, the right-handed top $t_R^c$ and two techni-fermions transforming as a doublet $Q_L$ and a singlet $N_R$ such that they can form a bound state with the quantum numbers of the Higgs.~\footnote{Technically there is a second choice, however no coupling of the Higgs to the top can be generated in that case. The hypercharge assignments in Table~\ref{fundamental theory} remain, therefore, unique.} The fact that the two techni-fermions need to transform as conjugate representations of the TC group fixes $\mathcal{G}_{\rm TC} = \SU(3)$. In fact, like for QCD, it is the only group for which a two-index antisymmetric representation coincides with the conjugate fundamental. The massless 28-plet baryon $\mathcal{B}_A$ thus consists of the following components
\beq
\mathcal{B}_A = \langle \xi_S \xi_{\bar{F}}^i \xi_{\bar{F}}^j \rangle \equiv
 \begin{pmatrix}
t_R^c & X_L & q_L \\
- X_L & N_R^c  & Q_L \\
-q_L & -Q_L & \tau_R^c
\end{pmatrix}\,,
\label{massless baryon}
\eeq
where $i,j = 1 \dots 8$ span over the 8 component of the global $\SU(8)$ symmetry, and the block form is based on the 3 $\xi_{\bar{F}}$ multiplets in Table~\ref{fundamental theory}.
Note that the doublet techni-fermion $Q_L$ has hypercharge $1/2$ while the singlet $N_R^c$ has hypercharge $0$, so that the TC bound state $\langle Q_L N_R^c \rangle$ transforms as a Higgs doublet. 
The model predicts the presence of two additional massless baryons: a singlet with hypercharge $1$ that can be identified with the SM right-handed tau, $\tau_R^c$, and a QCD-coloured techni-fermion with hypercharge $-1/3$, $X_R^c$. The presence of the latter suggests that the top mass will be generated in this model via partial compositeness~\cite{Kaplan:1991dc}. We stress the fact that this is a prediction of the model, as it was not taken in as a requirement for the choice of the matter content in Table~\ref{fundamental theory}.
The fermions in the bottom section of the table are added in order to cancel the gauge anomalies generated by the massless baryons in Eq.~\eqref{massless baryon}, and they include the right-handed bottom $b_R^c$ and the third-generation lepton doublet $l_L$. The complete theory remains chiral, while the TC theory after the GETC confinement is vector-like. Thus, it will condense and break the chiral symmetry {\it {\`a} la} QCD.

The mediators consist of the $\rho^\mu_{\rm GETC}$ mesons made by the $\xi_{\bar{F}}$ fermions, and they transform as the adjoint of the global $\SU(8)$ symmetry:
\beq
\rho^\mu_{\rm GETC} = \langle \bar{\xi}_{\bar{F}}^i \bar{\sigma}^\mu \xi_{\bar{F}}^i \rangle \equiv 
\begin{pmatrix}  G_c^\mu & C^\mu & D^\mu \\
C^{\dagger,\mu} & G_{\rm TC}^\mu & E^\mu \\
D^{\dagger,\mu} & E^{\dagger,\mu} & W_L^\mu \end{pmatrix} + \mbox{2 sing.}
\eeq
While the vectors on the diagonal transforms as the adjoints of the gauge groups, the off-diagonal ones generate the interesting couplings:
\begin{eqnarray}
& C^\mu = ({\tiny\yng(1)},\overline{\tiny\yng(1)},1)_{1/3}\,, \quad D^\mu = (1,\overline{\tiny\yng(1)},{\tiny\yng(1)})_{5/6}\,, & \nonumber \\
& E^\mu = (\overline{\tiny\yng(1)},1,{\tiny\yng(1)})_{1/2}\,, &
\end{eqnarray}
with $(\SU(3)_{\rm TC}, \SU(3)_c, \SU(2)_L)_{\UU(1)_Y}$ quantum numbers listed.
We define the $\rho^\mu_{\rm GETC}$ coupling to baryons as
\beq
\begin{split}
&g_\rho\; \bar{\mathcal{B}}_A \bar{\sigma}_\mu \left( \rho^\mu_{\rm GETC} \mathcal{B}_A + \mathcal{B}_A (\rho^\mu_{\rm GETC})^T \right) \\
=  &C_\mu\ J_C^\mu + D_\mu\ J_D^\mu + E_\mu\ J_E^\mu + \dots\,,  
\end{split}
\eeq
with currents given by
\begin{eqnarray}
\sqrt{2}\ J_C^\mu &=& \overline{t_R^c} \bar{\sigma}^\mu X_L + \overline{X_L} \bar{\sigma}^\mu N_R^c + \overline{q_L} \bar{\sigma}^\mu Q_L\,,\\
\sqrt{2}\ J_D^\mu &=& \overline{t_R^c} \bar{\sigma}^\mu q_L - \overline{X_L} \bar{\sigma}^\mu Q_L +  \overline{q_L} \bar{\sigma}^\mu  \tau_R^c\,, \\ 
\sqrt{2}\ J_E^\mu &=& \overline{X_L} \bar{\sigma}^\mu q_L + \overline{N_R^c} \bar{\sigma}^\mu Q_L + \overline{Q_L} \bar{\sigma}^\mu \tau_R^c\,.
\end{eqnarray}
Integrating out $C_\mu$, among the many four-fermion interactions we can identify the two below that induce partial compositeness for the top fields:
\beq
 \frac{g_\rho^2}{2 M_{\rho_C}^2}\  \left[ \left(  \overline{t_R^c} \bar{\sigma}^\mu X_L \right) \left( \overline{N_R^c}   \bar{\sigma}_\mu X_L \right) + \left( \overline{q_L} \bar{\sigma}^\mu Q_L \right) \left( \overline{N_R^c}   \bar{\sigma}_\mu X_L \right) \right]\,.
\eeq
%
Upon condensation of the TC interactions below $f_{\rm GETC}$, the couplings above will induce a mixing of the right and left-handed tops with the following techni-baryons: $T_R = \langle  X_L X_L \overline{N_R^c} \rangle$ and $T_L = \langle   X_L Q_L \overline{N_R^c}  \rangle$.
Similarly, integrating out $E_\mu$ generates the following coupling for $q_L$ and $\tau_R$:
\beq
 \frac{g_\rho^2}{2 M_{\rho_E}^2}\ \left[ \left( \overline{q_L}  \bar{\sigma}^\mu X_L \right)   \left(  \overline{N_R^c} \bar{\sigma}_\mu Q_L \right)  +   \left( \overline{\tau_R^c}  \bar{\sigma}^\mu   Q_L  \right)  \left(  \overline{N_R^c} \bar{\sigma}_\mu Q_L \right) \right]\,.
\eeq

Other four-fermion interactions generated by the above currents either contribute to couplings of the techni-$\rho$'s to SM fermions, or involve only techni-fermions (thus contributing to the potential for the techni-pNGBs).  
The meson $D_\mu$ is the only one that generates a coupling between SM fermions only, in the form:
\beq
 \frac{g_\rho^2}{2 M_{\rho_D}^2}\   \left(  \overline{t_R^c} \bar{\sigma}^\mu q_L \right) \left( \overline{\tau_R^c}  \bar{\sigma}_\mu  q_L \right)\,.
\eeq
This operator violates both B and L by one unit, while preserving $\rm B-L$, and contributes to the decay of the proton/neutron thus imposing a significant bound on the GETC scale. The most conservative constraint comes from $W$-loops that change the heavy flavours into light ones, leading to the dominant constraint from the neutron decay $n\to \bar{\nu} \pi^0$~\cite{Hou:2005iu}. Taking into account the improvement by a factor of 10 in the $n\to \nu \pi_0$ lifetime limit~\cite{Tanabashi:2018oca} and the fact that Ref.~\cite{Hou:2005iu} does not explicitly consider the chirality structure in our operator, we can estimate the limit on the coefficient of the operator to lie in the range $\left[10^{-14} ,10^{-11}\right]~\rm{TeV}^{-2}$, thus giving a lower bound on the GETC scale 
\beq
f_{\rm GETC} \approx \frac{M_\rho}{g_\rho} \gtrsim 10^{6\div 7}~\rm{TeV}\,.
\eeq
Referring to Eq.~\eqref{eq:ytop}, this implies that a walking phase between the GETC and TC condensation scales is necessary in this model.

The fundamental model described above suffers from many global $\UU(1)$'s that are not broken by the SM+TC gauging. This is a two-fold problem: the TC condensate spontaneously breaks some of them thus generating unwanted massless Goldstones; a coupling between the top fields and the composite Higgs is forbidden.
To amend this situation, we added to the model all the possible four-fermion interactions built out of the fermions in Table~\ref{fundamental theory}, assuming that they are generated at a higher scale $\Lambda_{\rm UV} \gg f_{\rm GETC}$.~\footnote{Note that some additional UV physics is also needed by the light generation masses.}
The new couplings can be classified according to the number of $\xi$'s contained: ones with 3 $\xi$'s generate effective masses for the 3 techni-fermions $X$, $Q$ and $N$ and explicitly break the $\UU(1)$'s already broken by the TC condensate thus avoiding the massless Goldstones; ones with 2 $\xi$'s have the same structure of the $\rho_{\rm GETC}$ couplings and thus do not break any $\UU(1)$.
Finally, there exist 3 operators that do not involve any $\xi$'s:
\begin{eqnarray}
&&\frac{c_{lb}}{\Lambda_{\rm UV}^2} (\overline{l_L} \bar{\sigma}^\mu b_R^c) (\overline{X_R^c} \bar{\sigma}_\mu Q_R^c)\,, \quad \frac{c_{\nu b}}{\Lambda_{\rm UV}^2} (\overline{N_L} \bar{\sigma}^\mu b_R^c) (\overline{X_R^c} \bar{\sigma}_\mu \nu_R^c)\,, \nonumber \\
&&\frac{c_{l\nu}}{\Lambda_{\rm UV}^2} (\overline{N_L} \bar{\sigma}^\mu l_L) (\overline{Q_R^c} \bar{\sigma}_\mu \nu_R^c)\,. \label{eq:leptoquarks}
\end{eqnarray}
Their effect at low energy is to generate lepto-quark couplings for some of the techni-mesons. They break all $\UU(1)$'s except two: one identifiable with the gauged $\UU(1)_Y$, and one matching the $\rm{B-L}$ charges on the SM fields, with charges defined in the last column of Table~\ref{fundamental theory}. With this added ingredients, the theory allows for a top Yukawa without any massless Goldstone.

We now briefly analyse the low energy end of the theory below the GETC confining scale, where it is the TC interactions that condense and give rise to a composite Higgs model. 
The TC strong sector has a global symmetry $\SU(6)_L \times \SU(6)_R$ acting on the massless techni-fermions
 \beq
 \psi_L = \begin{pmatrix}
 X_L \\
 Q_L \\
 N_L \end{pmatrix}\,, \qquad \psi_R^c = \begin{pmatrix}
 X_R^c \\
 Q_R^c \\
 N_R^c \end{pmatrix}\,,
 \eeq
which is broken down to the diagonal $\SU(6)$ by the condensate $\langle \psi_L \psi_R^c \rangle \neq 0$. The resulting pNGBs transform as the adjoint of $\SU(6)$, which decomposes under the SM gauge symmetries $(\SU(3)_c, \SU(2)_L)_{\UU(1)_Y}$ as
\begin{multline}
\pi_{\rm Adj} = (8,1)_0 \oplus (1,3)_0 \oplus 2\times (1,1)_0 \oplus \\
(1,2)_{1/2} \oplus (3,2)_{1/6} \oplus (3,1)_{-1/3} \oplus \mbox{h.c.}
\end{multline}
Besides the Higgs doublet, the light composite scalar spectrum contains two coloured ones matching the quantum numbers of a squark doublet and a sbottom singlet in supersymmetry (they decay like lepto-quarks via the couplings in Eq.~\eqref{eq:leptoquarks}), plus scalar partners of the SM gauge bosons. 
Note that this model does not contain a custodially invariant Higgs sector, fact linked to the common origin of the tops and techni-fermions. The TC theory  contains only 3 techni-fermions that are not charged under QCD, while the minimal custodial model requires 4~\cite{Ma:2015gra}.
However, the model can be easily extended by adding more elementary techni-fermions in vector-like pairs: a custodial model would be recovered by adding one singlet $E$ with hypercharge $-1$. This state also allows to write a four-fermion interaction to generate partial compositeness for the tau. Analogously, one can add a colour triplet to generate partial compositeness for the bottom quark. The extended model thus contains the following four Dirac elementary techni-fermions:
\beq
E = ({\tiny\yng(1)},1,1)_{-1}\,, \quad  X_b = ({\tiny\yng(1)},{\tiny\yng(1)},1)_{-2/3}\,.
\eeq
Remarkably, the TC sector of the extended model features 10 Dirac flavours under $\SU(3)_{\rm TC}$, which is believed to have an IR fixed point~\cite{Chiu:2016uui,Hasenfratz:2017qyr,Chiu:2018edw}, thus featuring the required walking between the GETC and the TC scales.~\footnote{For a study with opposed conclusions, see Ref.~\cite{Fodor:2018tdg}.}  The mass gap in the low energy theory can be generated by integrating out $X_b$, which needs to have a mass $m_{X_b} \approx \Lambda_{\rm TC}$, leaving only 7 light flavours in the low energy theory~\cite{Vecchi:2015fma}.~\footnote{The mechanism of generating a mass gap by giving mass to some techni-fermions in a theory with an IR conformal fixed point has been tested on the Lattice~\cite{Brower:2015owo,Hasenfratz:2016gut,Hasenfratz:2016uar} for $\SU(3)_{\rm TC}$.} This setting is technically natural, as  $m_{X_b}$ is a fundamental fermion mass, therefore it is radiatively stable.

Another interesting property of the model in Table~\ref{fundamental theory} is that it can be naturally embedded in a unified $\SU(5)$ model for the SM gauge interactions. Under the remaining $(\SU(5)_{\rm SM}, \SU(4)_{\rm GETC}, \SU(3)_{\rm TC})$, we have listed the fermion contents in Table~\ref{unified}.

\begin{table}[h]
\begin{tabular}{|c|c|c|c|}
\hline
& $SU(5)_{SM}$ & $SU(4)_{GETC}$ & $SU(3)_{TC}$  \\
\hline
$\left(\begin{matrix} \xi^1_{\bar{F}} \\ \xi^3_{\bar{F}}\end{matrix} \right)$ &$\tiny\yng(1)$&$\overline{\tiny\yng(1)}$&$1$\\
$\left(\begin{matrix} b^c_R \\ l_L\end{matrix} \right) $&$\overline{\tiny\yng(1)}$& $1$&$1$\\
$\left(\begin{matrix} X^c_R \\ Q^c_R\end{matrix} \right) $&$\overline{\tiny\yng(1)}$& $1$&$\overline{\tiny\yng(1)}$ \\
$\xi_s $&$1$& $\tiny\yng(2)$&$1$\\
$N_L $&$1$& $1$&$\tiny\yng(1)$ \\
$\xi^2_{\bar{F}}$&$1$&
$\overline{\tiny\yng(1)}$&$\tiny\yng(1)$ \\
\hline
\end{tabular}
 \caption{Fermion content of the minimal GETC model embedded in a unified $\SU(5)_{\rm SM}$ gauge interactions.}
 \label{unified}
\end{table}

In conclusion, in this letter we have shown a first simple model based on the {\it Global Extended Technicolor mechanism}, where a confining chiral theory at high energies can generate the top fields and the key ingredients for a composite Goldstone Higgs model with top partial compositeness at lower energies. Our model predicts the existence of a confining $\SU(3)$ theory with 6 light Dirac flavours in the fundamental (promoted to 7 if a custodial symmetry is required). Above $\Lambda_{\rm TC}$, we add 3 heavy flavours, so that the intermediate $\SU(3)$ model with 10 Dirac flavours can stay in a walking phase and explain the hierarchy between the TC and the GETC scales. This structure of the model derives from simple considerations on the GETC theory, and Lattice data will be needed in order to check the consistency of the model in terms of reproducing the top mass via large anomalous dimensions.
Our work thus encourages the study of $\SU(3)$ theories with 6 light and 4 heavy Dirac flavours in the fundamental representation. 
The low energy theory also features light pNGB scalars with the quantum numbers of the quark doublet and the right-handed bottom, which will decay like lepto-quarks as the theory only preserves $\rm B-L$.

\bigskip
\section*{Acknowledgements}
We acknowledge useful discussions with R.Contino and F.Sannino. 
We are grateful to the Mainz Institute for Theoretical Physics (MITP) of the DFG
Cluster of Excellence PRISMA+ (Project ID 39083149) for its hospitality and support
during the completion of this work.
GC and SV received partial support from the Labex-LIO (Lyon Institute of Origins) under grant ANR-10-LABX-66 (Agence Nationale pour la Recherche), and FRAMA (FR3127, F\'ed\'eration de Recherche ``Andr\'e Marie Amp\`ere''). Z-W.~W. is partially supported by the Danish National Research Foundation under the grant DNRF:90.

\bibliographystyle{JHEP-2-2}

\bibliography{mGETC.bib}

\end{document}